\documentclass[useAMS, usenatbib, fleqn]{mn2e}
\pdfoutput=1
\usepackage{aecompl}
\usepackage{amsmath}
\usepackage{amssymb}
\usepackage{color}
\usepackage{graphicx}
\usepackage{times}
\usepackage{url}
\usepackage{hyperref}
\hypersetup{colorlinks=true,citecolor=blue,linkcolor=blue,filecolor=blue,urlcolor=blue}
\usepackage{etoolbox}
\makeatletter
\patchcmd\@combinedblfloats{\box\@outputbox}{\unvbox\@outputbox}{}{%
   \errmessage{\noexpand\@combinedblfloats could not be patched}%
}%
 \makeatother

\newcommand*\diff{\mathop{}\!\mathrm{d}}
\renewcommand\vec{\mathbf}
\newcommand\vechat[1]{\mathbf{\hat #1}}
\newcommand\vectilde[1]{\mathbf{\tilde #1}}

\title[Dust-radiation coupling]
    {Simulating dust grain-radiation coupling on a moving mesh}
\author[R.~McKinnon et al.]
    {Ryan McKinnon,$^{1}$
     Rahul Kannan,$^{2}$\thanks{rahul.kannan@cfa.harvard.edu}
     Mark Vogelsberger,$^{1}$
     Stephanie O'Neil,$^{1}$
\newauthor
      Paul Torrey,$^{3}$ and Hui Li$^1$
     \vspace{0.3cm}\\
     $^{1}$Department of Physics and Kavli Institute for Astrophysics and Space Research,
           Massachusetts Institute of Technology,
           Cambridge, MA 02139, USA \\
     $^{2}$Center for Astrophysics $|$ Harvard $\&$ Smithsonian,
           Cambridge, MA 02138, USA \\
     $^{3}$Department of Astronomy,
           University of Florida,
           Gainesville, FL 32611, USA
    }
\begin{document}

\date{Accepted ???. Received ???; in original form ???}

\pagerange{\pageref{firstpage}--\pageref{lastpage}}
\pubyear{2018}

\maketitle

\label{firstpage}

\begin{abstract}
We present a model for the interaction between dust and radiation fields in the
radiation hydrodynamic code \textsc{arepo-rt}, which solves the moment-based
radiative transfer equations on an unstructured moving mesh.  Dust is directly
treated using live simulation particles, each of which represent a population of grains that are coupled to hydrodynamic motion through a drag force. We introduce
methods to calculate radiation pressure on and photon absorption by dust
grains.  By including a direct treatment of dust, we are able to
calculate dust opacities and update radiation fields self-consistently based on
the local dust distribution.  This hybrid scheme coupling dust particles to an unstructured mesh for radiation is validated using several test problems with
known analytic solutions, including dust driven via spherically-symmetric flux
from a constant luminosity source and photon absorption from radiation incident
on a thin layer of dust.  Our methods are compatible with the multifrequency
scheme in \textsc{arepo-rt}, which treats UV and optical photons as
single-scattered and IR photons as multi-scattered.  At IR wavelengths, we
model heating of and thermal emission from dust.  Dust and gas are not assumed
to be in local thermodynamic equilibrium but transfer energy through
collisional exchange.  We estimate dust temperatures by balancing these
dust-radiation and dust-gas energy exchange rates.  This framework for coupling
dust and radiation can be applied in future radiation hydrodynamic simulations
of galaxy formation.
\end{abstract}


\section{Introduction}

In the interstellar medium, dust grains and radiation fields affect one
another.  Grains driven by radiation pressure can transfer momentum to nearby
gas, helping to regulate star formation and launch galactic outflows
\citep{Krumholz2009, Murray2010, Hopkins2012}.  Dust scatters and absorbs
starlight and emits in the infrared \citep[IR;][]{Schlegel1998, Bernstein2002,
Schlafly2011}, influencing galactic spectral energy distributions
\citep{Silva1998, Dale2001}.  The temperatures of dust grains are affected by
radiative heating, thermal emission, and collisional energy exchange with gas
\citep{Hollenbach1979, Dwek1986, Boulanger1988}, and small grains in particular
are susceptible to non-equilibrium temperature fluctuations \citep{Dwek1986,
Guhathakurta1989, Siebenmorgen1992}.  Galaxy formation models must consider
these physical processes when accounting for the impact of dust and radiation.

Owing to the computational cost, many works do not directly treat dust or
radiation.  Instead, they attempt to capture the effects of radiation pressure
on dust grains through physically-motivated subgrid methods \citep{Murray2005,
Hopkins2011, Agertz2013}.  This typically involves injecting momentum or
thermal energy in gas surrounding sources of radiation like stars, avoiding the
need to evolve radiation fields.  For example, to mimic the impact of radiation
multi-scattered by dust, momentum injection rates can increase as a function of
local IR optical depth.  Such radiation pressure feedback models have been used
in cosmological settings and can reduce star formation \citep{Aumer2013, Hopkins2014, Roskar2014, Agertz2015}.  To improve the accuracy of
these subgrid radiation pressure models, there have been renewed efforts to
develop numerical methods suited for simulations with limited resolution
\citep{Krumholz2018, Hopkins2019}.

In recent years, a variety of methods have been developed to more directly
model radiation and its effect on dust.  For example, Monte Carlo methods track
the emission and absorption of individual photon packets but can be
computationally expensive to run \citep[e.g.][]{Bjorkman2001, Oxley2003,
Camps2015b, Tsang2015, Smith2019}.  Other methods use long-characteristic ray
tracing to solve the exact radiative transfer equation \citep[e.g.][]{Abel1999,
Abel2002, Wise2012, Greif2014, Jaura2018}, although this can scale unfavourably
with the number of sources.  Alternatively, some mesh-based radiation
hydrodynamics solvers combine moments of the radiative transfer equation with
the M1 closure relation \citep{Levermore1984, Dubroca1999}, in which the
Eddington tensor is calculated strictly from local quantities and is
independent of the number of sources.  Various codes employ the M1 closure
\citep{Gonzalez2007, Rosdahl2013, Rosdahl2015b, Kannan2019a} and have been used
in radiation hydrodynamic simulations of isolated disc galaxies
\citep{Rosdahl2015, Kannan2019c},  quasar outflows via radiation pressure \citep{Bieri2017, Costa2018b, Barnes2018}, and reionisation calculations \citep[see for eg. ][]{sphinx, Wu2019a, Wu2019b}.

In this work, we evolve radiation using the radiation hydrodynamics solver
\textsc{arepo-rt} \citep{Kannan2019a} integrated into the unstructured
moving-mesh code \textsc{arepo} \citep{Springel2010}.  \citet{Kannan2019a}
includes a simplified model for radiation pressure on, photon absorption by,
and IR emission from dust grains, treating dust as a passive scalar perfectly
coupled to hydrodynamic motion and in local thermodynamic equilibrium with gas.
In this work, we relax those assumptions and adopt a more general treatment of
dust in the context of radiation hydrodynamics.  We do not treat dust as an
element of gas cells but instead using simulation particles
\citep{McKinnon2018}.  This approach allows richer and more realistic dynamical
and thermal interactions among dust, gas, and radiation.  We present our
methods and test problems in Section~\ref{SEC:methods} and conclude in
Section~\ref{SEC:conclusion}.

\section{Methods}\label{SEC:methods}

We start with a brief summary of the equation of motions for dust and gas components coupled through
aerodynamic drag and possibly subject to external accelerations.  We refer the
reader to Section~2 in \citet{McKinnon2018} for background details.  A dust
grain of mass $m_\text{d}$ feels an acceleration
\begin{equation}
\frac{\diff \vec{v}_\text{d}}{\diff t} = -\frac{K_\text{s} (\vec{v}_\text{d} - \vec{v}_\text{g})}{m_\text{d}} + \vec{a}_\text{d,ext},
\label{EQN:dvdt_dust}
\end{equation}
where $\vec{v}_\text{d}$ and $\vec{v}_\text{g}$ are the local dust and gas
velocities, respectively, $\vec{a}_\text{d,ext}$ accounts for additional
sources of dust acceleration (e.g.~gravity or radiation pressure), and $K_\text{s}$ is a drag coefficient.  Similarly, the gas acceleration is given by
\begin{equation}
\frac{\diff \vec{v}_\text{g}}{\diff t} = -\frac{\nabla P}{\rho_\text{g}} + \frac{\rho_\text{d} K_\text{s} (\vec{v}_\text{d} - \vec{v}_\text{g})}{\rho_\text{g} m_\text{d}} + \vec{a}_\text{g,ext},
\label{EQN:dvdt_gas}
\end{equation}
where $P$ denotes gas pressure, $\rho_\text{d}$ and $\rho_\text{g}$ are the
dust and gas density, respectively, and $\vec{a}_\text{g,ext}$ is the external
gas acceleration. For use later, we define the dust-to-gas ratio $D \equiv
\rho_\text{d} / \rho_\text{g}$.  Here, the drag backreaction force by dust on
gas is exactly opposite to the drag force by gas on dust.  For drag in the
Epstein regime, the equations of motion can be rewritten in terms of
$t_\text{s} \equiv m_\text{d} / [K_\text{s} (1 + D)]$, the stopping time-scale
for drag, whose value is given by the approximation
\begin{equation}
t_\text{s} = \frac{\sqrt{\pi \gamma} a \rho_\text{gr}}{2 \sqrt{2} \rho c_\text{s}} \left( 1 + \frac{9 \pi}{128} \left|\frac{\vec{v}_\text{d} - \vec{v}_\text{g}}{c_\text{s}} \right|^2 \right)^{-1/2},
\label{EQN:t_s}
\end{equation}
for a given grain size $a$, internal solid density of dust grains
$\rho_\text{gr}$, total density $\rho \equiv \rho_\text{g} + \rho_\text{d}$,
and gas sound speed $c_\text{s}$ \citep{Paardekooper2006, Price2017}.  The
dust-to-gas ratio $D$ affects the backreaction of dust dragging gas: when $D
\ll 1$, the gas drag acceleration is significantly smaller than the dust
acceleration.  Since $D \approx 0.01$ for the Milky Way and other nearby
galaxies \citep[e.g.][]{Draine2007}, in galaxy simulations it is often
reasonable to neglect the backreaction of drag on gas.

However, dust dynamics are applicable in a wide variety of settings, and it is
worthwhile to relax the assumption of low dust-to-gas ratio.  Below, we extend
the drag implementation of \citet{McKinnon2018}, which neglected drag
backreaction on gas, to handle drag coupling at arbitrary dust-to-gas ratio.
We adopt the second-order semi-implicit time integrator presented in
\citet{LorenAguilar2015}.  Defining the functions
\begin{equation}
\xi(\Delta t) = \frac{1 - e^{- \Delta t / t_\text{s}}}{1 + D}
\end{equation}
and
\begin{equation}
\Lambda(\Delta t) = (\Delta t + t_\text{s}) \xi(\Delta t) - \frac{\Delta t}{1 + D},
\end{equation}
the velocity update of a dust particle from time $t$ to $t + \Delta t$ is given
by
\begin{equation}
\begin{split}
\vec{v}_\text{d}(t + \Delta t) &= \vectilde{v}_\text{d}(t + \Delta t) - \xi(\Delta t) \left[\vectilde{v}_\text{d}(t + \Delta t) - \vectilde{v}_\text{g}(t + \Delta t)\right] \\
&\quad + \Lambda(\Delta t) \left[\vec{a}_\text{d,ext}(t) - \vec{a}_\text{g,ext}(t) + \frac{\nabla P}{\rho_\text{g} }\right],
\end{split}
\label{EQN:dust_velocity_update}
\end{equation}
where $\vec{v}_\text{d}(t + \Delta t)$ is the dust particle velocity after the
drag update and $\vectilde{v}_\text{d}(t + \Delta t)$ and
$\vectilde{v}_\text{g}(t + \Delta t)$ are the dust and gas velocities at time
$t + \Delta t$ after accounting for non-drag accelerations.  We apply
equation~(\ref{EQN:dust_velocity_update}) in the following SPH-like fashion to
conserve total momentum.  After calculating the change in a dust particle's
momentum $\Delta \vec{p}_\text{d}$, we subtract momentum $w_i \Delta
\vec{p}_\text{d}$ from each of the $N_\text{ngb}$ closest gas cells, where
$w_i$ is a weight assigned to cell $i$ in this set.  The amount of momentum
change in each neighboring gas cell is kernel-weighted so that closer gas cells
lose a greater fraction of $\Delta \vec{p}_\text{d}$.  In our work, we adopt
the standard cubic spline kernel.

The dissipation of total dust and gas kinetic energy from drag leads to
frictional heating of the gas \citep[e.g.][]{Laibe2012}.  When applying
equation~(\ref{EQN:dust_velocity_update}), we calculate the change in kinetic
energy of the dust particle $\Delta K_\text{d}$.  Then, after updating the
momentum of local gas cell $i$, we determine its change in kinetic energy
$\Delta K_\text{g}$.  The internal energy of the cell is then increased by
$-(w_i \Delta K_\text{d} + \Delta K_\text{g})$.  Summing over all cells about a
dust particle, the gas internal energy increases by the amount total dust and
gas kinetic energy decreases.

\subsection{Interaction between dust grains and radiation}

Gravity and drag are not the only dynamical forces that act on interstellar
dust grains.  Since dust provides a source of opacity to radiation, the
resulting radiation pressure force \citep[e.g.][]{Weingartner2001c} can also
affect dust and gas dynamics.  This has potential implications for the
efficiency of stellar feedback and galactic outflows.  For example, some models
predict that multi-scattering of IR radiation due to dust opacity can enhance
self-regulation of star formation \citep{Murray2010, Hopkins2011, Agertz2013}
and boost gas outflows in galaxies up to IR optical depths of $\tau_\text{IR}
\approx 10$ \citep{Thompson2015, Costa2018}.  However, others suggest that dust
reprocessing of radiation does not strongly affect wind momentum flux
\citep{Krumholz2013, Reissl2018}.  The interaction between dust and radiation can also
affect the CGM in addition to the ISM.  Radiation pressure can efficiently
drive dust grains into galactic haloes, with grains of different chemical
composition feeling different strength forces \citep{Ferrara1991}.  Given the
reliance of galaxy formation simulations on feedback physics, it is natural to
want to model the interaction between dust and radiation in a more direct
manner.

Recently, \citet{Kannan2019a} developed radiation hydrodynamics methods for the
moving-mesh code \textsc{arepo} \citep{Springel2010}, capable of tracking
multifrequency radiation transport and thermochemistry.  Here, we briefly
summarise these methods and refer the reader to Sections~2 and~3 of
\citet{Kannan2019a} for a more complete description.  In general, radiative transfer is
formulated in terms of specific intensity $I_\nu(\vec{x}, \vec{n}, t)$, which
at position $\vec{x}$ and time $t$ quantifies the rate of change of radiation
energy at frequency $\nu$ per unit time, per unit area in the direction
$\vec{n}$, per unit solid angle, and per unit frequency.  This evolves
according to the radiative transfer equation,
\begin{equation}
\frac{1}{\tilde{c}} \frac{\partial I_\nu}{\partial t} + \vec{n} \cdot \nabla I_\nu = j_\nu - \kappa_\nu \rho I_\nu,
\end{equation}
where $\tilde{c}$ is the propagation speed of radiation, $j_\nu$ denotes the
emission coefficient, $\kappa_\nu$ is the gas opacity, and $\rho$ is the gas
density.  Numerically solving the radiative transfer equation is challenging,
given the large number of variables.  One approach is to take moments of the
radiative transfer equation, yielding
\begin{equation}
\frac{\partial E}{\partial t} + \nabla \cdot \vec{F} = S - \kappa_\text{E} \rho \tilde{c} E
\label{EQN:E_moment}
\end{equation}
and
\begin{equation}
\frac{\partial F}{\partial t} + \tilde{c}^2 \nabla \cdot \mathbb{P} = -\kappa_\text{F} \rho \tilde{c} \vec{F},
\label{EQN:F_moment}
\end{equation}
where $S$ denotes the emission coefficient integrated over frequency and solid
angle, $\kappa_\text{E}$ and $\kappa_\text{F}$ are the energy- and
flux-weighted mean opacities, and the radiation energy density $E$, flux $F$,
and pressure $\mathbb{P}$ are given by
\begin{equation}
\{ \tilde{c} E, F, \mathbb{P} \} = \int_{\nu_1}^{\nu_2} \int_{4\pi} \{ 1, \vec{n}, \vec{n} \otimes \vec{n} \} I_\nu \diff \Omega \diff \nu.
\end{equation}
All quantities are integrated over a frequency interval $[\nu_1, \nu_2]$.  To
simplify this system, \textsc{arepo-rt} employs the M1 closure
\citep{Levermore1984, Dubroca1999} that calculates the pressure strictly in
terms of the local radiation energy density and flux.  This approximation
yields a computational cost independent of the number of radiation sources.
Equations~(\ref{EQN:E_moment}) and~(\ref{EQN:F_moment}) are solved on an
unstructured moving mesh using an operator-split approach, breaking the moment
equations into transport and source and sink terms.  To calculate intercell
fluxes, \textsc{arepo-rt} employs a Harten-Lax-van Leer or global
Lax-Friedrichs Riemann solver.

The propagation of radiation also affects hydrodynamics.  The momentum
conservation equation takes the form
\begin{equation}
\frac{\partial (\rho \vec{v})}{\partial t} + \nabla \cdot (\rho \vec{v} \otimes \vec{v} + P \mathbb{I}) = \frac{\kappa_\text{F} \rho F}{c},
\end{equation}
where $\vec{v}$ is the gas velocity, $P$ is the thermal pressure, $\mathbb{I}$
is the identity tensor, and $c$ is the speed of light.  In \citet{Kannan2019a},
the opacities entering the above equations are calculated for individual gas
cells using crude approximations or a simplified dust model treating dust as a
passive scalar perfectly coupled to hydrodynamic motion \citep{McKinnon2016}.
In this work, we instead model dust using live simulation particles
\citep{McKinnon2018} and not as a property of gas cells.  We describe below how
the methods in \citet{Kannan2019a} can be extended to model radiation pressure
on and photon absorption by dust grains using this hybrid gas cell and dust
particle scheme.  As a result, dust opacities are calculated self-consistently
using the local distribution of dust.  We generalize these methods to
multifrequency radiation and distributions of grain sizes and discuss how dust
grains can exchange energy with IR radiation and gas.

\subsubsection{Calculating radiation pressure on populations of grains}\label{SEC:Q_pr_sizes}

The methods described in \citet{McKinnon2018} model dust using simulation
particles representing ensembles of grains of different sizes.  Each particle
has a grain size distribution, discretised into $N_\text{GSD}$ grain size bins
with different typical sizes.  Similarly, the radiative transfer scheme
introduced in \citet{Kannan2019a} solves for the propagation of radiation in
$N_\text{RT}$ different frequency bins (e.g.~to model different UV wavelength
ranges or simultaneous UV and IR radiation fields).

We calculate the radiation pressure force $\vec{f_\text{pr}}$ on a dust
particle with grains of different sizes in a kernel-smoothed manner,
interpolating over $N_\text{ngb}$ neighboring gas cells.  In the general case
of multifrequency, anisotropic radiation \citep{Weingartner2001c}, this
radiation pressure force takes the form
\begin{equation}
\vec{f_\text{pr}} = \sum_{k = 0}^{N_\text{ngb} - 1} w_k \left[ \sum_{i = 0}^{N_\text{GSD} - 1} \sum_{j = 0}^{N_\text{RT} - 1} \left( \frac{\vec{F_{j,k}}}{c} \right) N_i \pi {a_i^\text{c}}^2 Q_\text{pr}(a_i^\text{c}, \lambda_j) \right],
\label{EQN:f_pr}
\end{equation}
where $\vec{F_{j,k}}$ and $\lambda_j$ are the energy flux and (mean) wavelength
in radiative transfer bin $j$ for neighboring gas cell $k$, and $N_i$ and
$a_i^\text{c}$ are the number of grains and midpoint of grain size bin $i$.  We
use the same cubic spline kernel as in \citet{McKinnon2018} to calculate the
weight $w_k$ for gas cell $k$ as a function of its distance from the given dust
particle.  These weights are normalised such that they sum to unity over all
neighboring gas cells.  In equation~(\ref{EQN:f_pr}), the term in brackets
represents what the radiation pressure force would be if all grain cross
section were concentrated in gas cell $k$, summing over all grain sizes and
frequency bins.

This force depends on the radiation pressure efficiency $Q_\text{pr}(a,
\lambda)$, which is a dimensionless factor denoting the ratio of radiation
pressure cross section to geometric cross section for a dust grain of size $a$
and wavelength of incident radiation $\lambda$.  This efficiency factor is
defined as $Q_\text{pr} \equiv Q_\text{abs} + (1 - \langle \cos \theta \rangle)
Q_\text{sca}$, where $Q_\text{abs}$ is the ratio of absorption cross section to
geometric cross section, $Q_\text{sca}$ is the ratio of scattering cross
section to geometric cross section, $\langle \cos \theta \rangle$ is the
average cosine of scattering angle $\theta$, and all three quantities are
functions of grain size and wavelength.

Following \citet{McKinnon2018}, we use tabulated values of $Q_\text{abs}$,
$Q_\text{sca}$, and $\langle \cos \theta \rangle$ for silicate and graphite
grains from \citet{Draine1984} and \citet{Laor1993}.  These tabulations cover
the grain size range $0.001 \, \mu\text{m} < a < 10 \, \mu\text{m}$ and
wavelength range $0.001 \, \mu\text{m} < \lambda < 1000 \, \mu\text{m}$.  We
perform two-dimensional logarithmic interpolation to calculate $Q_\text{pr}(a,
\lambda)$ values at arbitrary grain size and wavelength, averaging results for
silicate and graphite grains.

\subsubsection{Absorption of photons from dust grain opacity}

Dust grains absorb and reradiate in the IR a sizeable fraction of starlight,
often estimated around 30 per cent \citep{Soifer1991, Popescu2002}.  We first
describe absorption in the single-scattering regime \citep[e.g.~for UV
frequency bins, as in Section~3.2.1 of][]{Kannan2019a}. 

We account for photon absorption in a kernel-smoothed fashion, by spreading a
dust particle's absorption cross section among neighboring gas cells to update
energy densities and fluxes.  This approach effectively replaces the dust
opacity terms in equations~47 and~48 in \citet{Kannan2019a} that are used in
simulations lacking live dust particles.  Specifically, when processing an
active dust particle we update the energy density in radiation bin $j$ for
neighboring gas cell $k$ according to the rate
\begin{equation}
\frac{\partial E_{j,k}}{\partial t} = -\frac{w_k \tilde{c} E_{j,k}}{V_k} \sum_{i = 0}^{N_\text{GSD} - 1} N_i \pi {a_i^\text{c}}^2 Q_\text{abs}(a_i^\text{c}, \lambda_j),
\label{EQN:E_attenuation}
\end{equation}
where $w_k$ is the gas cell's kernel weight, $\tilde{c}$ is the reduced speed
of light, and $V_k$ is the gas cell volume.  As in equation~(\ref{EQN:f_pr}),
$N_i$ is the number of grains in size bin $i$ for the dust particle,
$a_i^\text{c}$ is the grain size at the midpoint of size bin $i$, and
$\lambda_j$ is the (mean) wavelength in radiation bin $j$.  The absorption
cross section efficiency $Q_\text{abs}$, described in
Section~\ref{SEC:Q_pr_sizes}, is a function of grain size and wavelength of
radiation.  Likewise, the flux in frequency bin $j$ of gas cell $k$ is updated
using the rate
\begin{equation}
\frac{\partial \vec{F_{j,k}}}{\partial t} = -\frac{w_k \tilde{c} \vec{F_{j,k}}}{V_k} \sum_{i = 0}^{N_\text{GSD} - 1} N_i \pi {a_i^\text{c}}^2 Q_\text{abs}(a_i^\text{c}, \lambda_j).
\label{EQN:F_attenuation}
\end{equation}

\subsubsection{Multifrequency radiation and dust coupling}\label{SEC:multifrequency_radiation}

The radiation hydrodynamics methods in \citet{Kannan2019a} are capable of treating multifrequency radiation, divided
into several UV and optical bins (e.g.~corresponding to H\,\textsc{i},
He\,\textsc{i}, and He\,\textsc{ii} ionisation) and one IR bin.  Photons in UV
and optical bins are ``single-scattered'': the energy absorbed by dust at these
wavelengths is reemitted as IR radiation.  On the other hand, IR photons are
``multi-scattered'': dust grains can absorb IR photons but also thermally emit
in the IR.

In the notation below, we assume that the $N_\text{RT}$ radiation bins are
ordered by decreasing frequency, so that the IR bin is last.  Then, the IR energy density in gas cell $k$, $E_{\text{IR},k}$, changes according to
\begin{equation}
\frac{\partial E_{\text{IR},k}}{\partial t} = \frac{\partial E_{\text{IR},k}^\text{reprocess}}{\partial t} + \frac{\partial E_{\text{IR},k}^\text{thermal}}{\partial t},
\end{equation}
the sum of reprocessing rates (i.e.~energy from UV and optical photons that is
reemitted in the IR) and thermal dust exchange rates (i.e.~dust emission and
absorption of IR photons), respectively.  In this section, we focus on
reprocessing of UV and optical radiation and discuss thermal IR exchange next
in Section~\ref{SEC:thermal_coupling}.  Then, the rate at which the IR energy
density increases is given by the total rate at which UV and optical energy
density decreases.  That is,
\begin{equation}
\frac{\partial E_{\text{IR},k}^\text{reprocess}}{\partial t} = - \sum_{j = 0}^{N_\text{RT}-2} \frac{\partial E_{j,k}}{\partial t}
\end{equation}
where the sum is over all non-IR bins and the energy absorption rate in
radiation bin $j$, $\partial E_{j,k} / \partial t$, is computed via
equation~(\ref{EQN:E_attenuation}).  Additionally, paralleling
equation~(\ref{EQN:F_attenuation}), the IR flux in gas cell $k$ evolves
according to
\begin{equation}
\frac{\partial \vec{F_{\text{IR},k}}}{\partial t} = -\frac{w_k \tilde{c} \vec{F_{\text{IR},k}}}{V_k} \sum_{i = 0}^{N_\text{GSD} - 1} N_i \pi {a_i^\text{c}}^2 Q_\text{abs}(a_i^\text{c}, \lambda_\text{IR}).
\end{equation}

\subsubsection{Thermal coupling for infrared radiation}\label{SEC:thermal_coupling}

Dust grains both emit and absorb IR photons \citep{Dwek1986, Krumholz2013} and
also collisionally exchange energy with gas \citep{Hollenbach1979, Omukai2000,
Goldsmith2001}.  These competing processes affect the temperature of a
population of dust grains \citep{Goldsmith2001, Krumholz2014, Smith2017}.
Small grains in particular are susceptible to stochastic heating and
temperature fluctuations \citep{Guhathakurta1989, Siebenmorgen1992,
Draine2001}, and methods to calculate and evolve dust temperature probability
distributions are used in various applications
\citep[e.g.][]{Pavlyuchenkov2012, Camps2015}.  However, given the limited
resolution our simulations, we instead follow the approach in
\textsc{despotic}~\citep{Krumholz2014} and \textsc{grackle}~\citep{Smith2017},
treating dust grains as being in thermal equilibrium and neglecting such
temperature fluctuations.

As with equation~(\ref{EQN:E_attenuation}), we perform the coupling between
dust, gas, and IR radiation in a kernel-smoothed manner, where dust particles
exchange energy with their surrounding neighbors.  Since dust particles are
superpositions of grains of different sizes, we allow for the possibility that
the dust temperature varies from one grain size bin to another.  When
considering the exchange between a dust particle and neighboring gas cell $k$,
the IR radiation energy density $E_{\text{IR},k}$, dust energy density
$u_\text{d}$, and gas energy density $u_{\text{g},k}$ evolve according to the
system
\begin{equation}
\begin{split}
&\frac{\partial E_{\text{IR},k}^\text{thermal}}{\partial t} = \Lambda_{\text{dr},k} \\
&\frac{\partial u_{\text{d}}}{\partial t} = -\Lambda_{\text{dr},k} - \Lambda_{\text{dg},k} \\
&\frac{\partial u_{\text{g},k}}{\partial t} = \Lambda_{\text{dg},k}.
\end{split}
\label{EQN:thermal_coupling_system}
\end{equation}
Here, the dust-radiation energy exchange rate per unit volume
\citep[e.g.][]{Krumholz2013} in gas cell $k$ is given by $\Lambda_{\text{dr},k}
= \sum_{i=0}^{N_\text{GSD}-1} \Lambda_{\text{dr},k,i}$, where the contribution
from grains in size bin $i$ is
\begin{equation}
\Lambda_{\text{dr},k,i} = \left[ \frac{w_k}{V_k} N_i \pi {a_i^\text{c}}^2 Q_\text{abs}(a_i^\text{c}, \lambda_\text{IR}) \right] (c a_\text{B} T_{\text{d},k,i}^4 - \tilde{c} E_{\text{IR},k}),
\label{EQN:Lambda_dr}
\end{equation}
where $w_k$ is the gas cell's kernel weight (see equation~\ref{EQN:f_pr}),
$V_k$ is the cell volume, $a_\text{B}$ is the radiation constant, and
$T_{\text{d},k,i}$ is the cell's dust temperature for grains in size bin $i$.
The prefactor in equation~(\ref{EQN:Lambda_dr}) is the cross section per unit
volume in gas cell $k$, obtained by assigning a weighted fraction of the dust
particle's grain population to the cell.  Also, the dust-gas energy exchange
rate per unit volume \citep[e.g.][]{Burke1983, Hollenbach1989} is calculated
via $\Lambda_{\text{dg},k} = \sum_{i=0}^{N_\text{GSD}-1}
\Lambda_{\text{dg},k,i}$, using the exchange rate with grains in size bin $i$
of
\begin{equation}
\Lambda_{\text{dg},k,i} = \left[ \frac{w_k}{V_k} N_i \pi {a_i^\text{c}}^2 \right] n_{\text{H},k} v_{\text{th},k} \overline{\alpha}_T (2 k_\text{B}) (T_{\text{d},k,i} - T_{\text{g},k}).
\end{equation}
Here, $n_{\text{H},k}$ is the hydrogen number density in gas cell $k$,
$\overline{\alpha}_T \approx 0.5$ is an estimated ``accommodation coefficient''
\citep[see equation~9 in][]{Burke1983}, $k_\text{B}$ is the Boltzmann constant,
$T_{\text{g},k}$ is the gas temperature, and
\begin{equation}
v_{\text{th},k} \equiv \left( \frac{8 k_\text{B} T_{\text{g},k}}{\pi m_\text{p}} \right)^{1/2}
\end{equation}
is the gas thermal velocity.

\begin{figure*}
\centering
\includegraphics{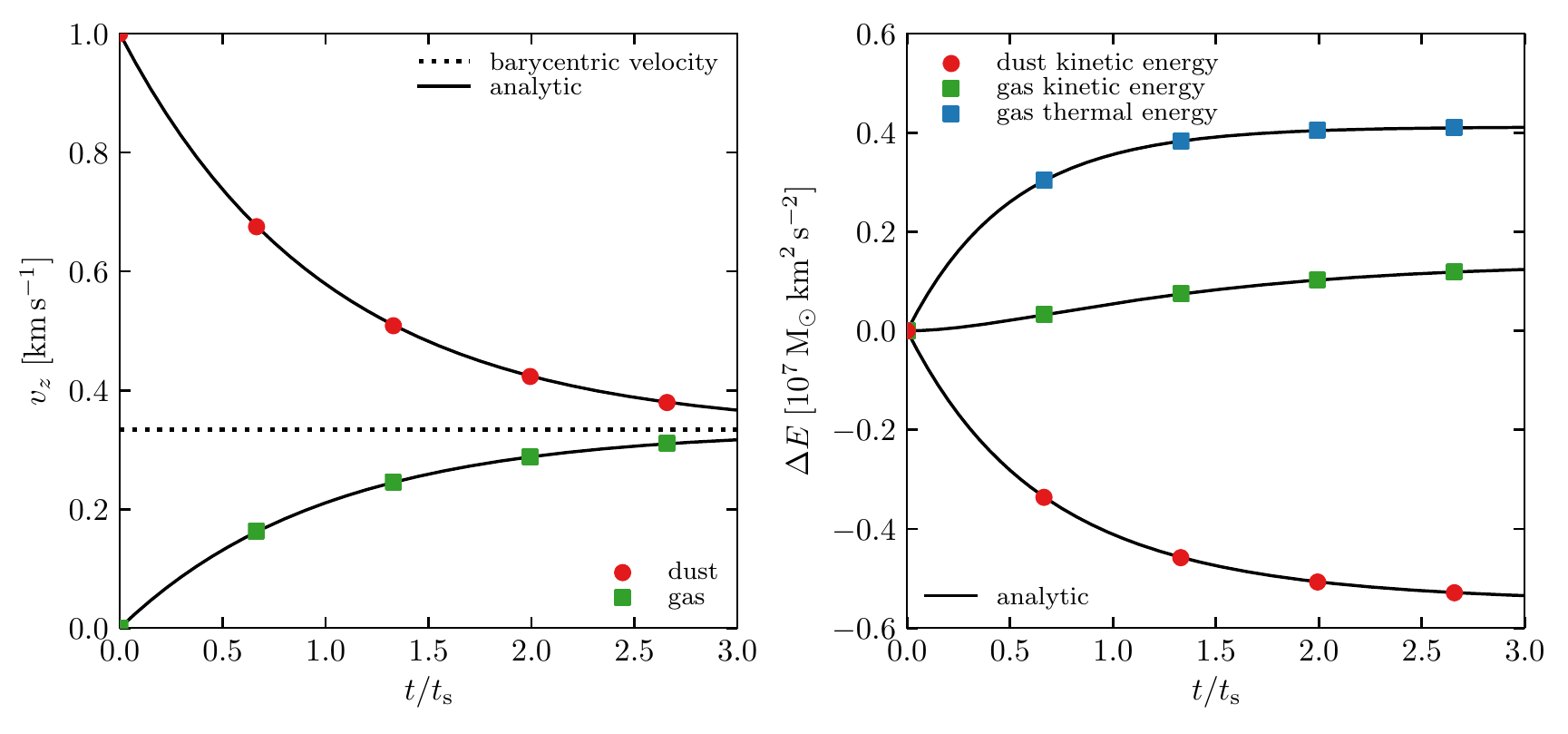}
\caption{Left panel: velocity profiles versus time for a dust and gas mixture
coupled via drag with an initial dust-to-gas ratio of $D = 0.5$.  Coloured
points show simulation results as the dust and gas velocities approach the
barycentric velocity (dotted line), while the solid lines indicate the analytic
predictions.  Time is given in units of the drag stopping time-scale
$t_\text{s}$.  Right panel: change in different energy components versus time.
Dust kinetic energy (red) decreases, while gas kinetic energy (green)
increases.  The net loss of kinetic energy from drag leads to an increase in
gas thermal energy (blue).}
\label{FIG:dustybox_backreaction_evolution}
\end{figure*}

Since dust grains reach thermal equilibrium on a rapid time-scale
\citep{Woitke2006, Krumholz2014, Smith2017}, we calculate the dust temperature
$T_{\text{d},k,i}$ that grains in size bin $i$ in gas cell $k$ have by solving
the instantaneous equilibrium condition
$\Lambda_{\text{dr},k,i}(T_{\text{d},k,i}) +
\Lambda_{\text{dg},k,i}(T_{\text{d},k,i}) = 0$ using Newton's method for
root-finding.  Using this dust temperature, over the dust particle's time-step
$\Delta t$ we then add $\Lambda_{\text{dr},k,i} V_k \Delta t$ in IR radiation
energy and $\Lambda_{\text{dg},k,i} V_k \Delta t$ in thermal energy to gas cell
$k$.  This process is repeated for all grain size bins and all gas cells within
a dust particle's smoothing length.  An average dust temperature for size bin
$i$ in the dust particle can be calculated as $T_{\text{d},i} =
\sum_{k=0}^{N_\text{ngb}-1} w_k T_{\text{d},k,i}$.

%
%

\section{Test Problems}\label{SEC:result}
\subsection{Demonstration of drag and kinetic energy dissipation}

When the dust-to-gas ratio $D$ is of order unity, drag impacts both dust and
gas.  To illustrate this in a simple test, we initialise a box of side length
$1 \, \text{kpc}$ with $32^3$ gas cells and $32^3$ dust particles arranged on
an equispaced lattice.  Gas cell masses are set so that the box has a uniform
gas density of $n = 1 \, \text{cm}^{-3}$.  We adopt a dust-to-gas ratio of $D =
0.5$, so that the mass of a dust particle is half the mass of a gas cell.
Initially, gas is at rest ($\vec{v}_\text{g}(t = 0) = \vec{0} \, \text{km} \,
\text{s}^{-1}$), and dust is given nonzero velocity along the $z$ axis
($\vec{v}_\text{d}(t = 0) \equiv \vec{v}_{0} = 1 \, \text{km} \, \text{s}^{-1}
\, \vechat{z}$).

Dust particles are assumed to consist of $0.1 \, \mu\text{m}$ grains, and the
internal energy per unit mass of the gas is initially $u = 1000 \,
\text{km}^{2} \, \text{s}^{-2}$.  From the values above, one can calculate the
stopping time-scale $t_\text{s}$ over which the drag force acts.  In this test,
the analytic prediction for gas velocity is
\begin{equation}
\vec{v}_\text{g}(t) = \left( \frac{D}{1 + D} \right) \left( 1 - e^{-t/t_\text{s}} \right) \vec{v}_{0},
\end{equation}
and the analytic prediction for dust velocity is
\begin{equation}
\vec{v}_\text{d}(t) = \left( \frac{1}{1 + D} \right) \left( D + e^{-t/t_\text{s}} \right) \vec{v}_{0}.
\end{equation}
As $t \to \infty$, both $\vec{v}_\text{g}$ and $\vec{v}_\text{d} \to D
\vec{v}_{0} / (1 + D) = \vec{v}_{0} / 3$, showing that gas and dust are
expected to move at the barycentric velocity of the mixture.

Figure~\ref{FIG:dustybox_backreaction_evolution} shows the time evolution of
dust and gas velocities and energies for this mixture coupled by drag.  Since
the drag force acts in an equal and opposite manner on dust and gas components
and the dust-to-gas ratio is less than unity, dust velocities decrease more
quickly than gas velocities increase.  The initial relative velocity between
dust and gas rapidly decays.  After just two stopping time-scales, the dust and
gas velocities are already within thirty per cent of the barycentric velocity.
Over time, more dust kinetic energy is lost than gas kinetic energy is gained.
The net dissipation of kinetic energy leads to an increase in gas thermal
energy.  In both panels of Figure~\ref{FIG:dustybox_backreaction_evolution},
simulation results for velocity and energy evolution are visually
indistinguishable from analytic predictions.

\subsection{Spherically symmetric radiation pressure test problem}

To demonstrate our radiation pressure scheme, we first consider a monochromatic
radiation source with constant luminosity $L$ and grains of a single fixed size
$a$.  Analytically, a dust grain at distance $r$ from the radiation source
feels a radial force with magnitude
\begin{equation}
f_\text{pr} = \left( \frac{L}{4 \pi r^2 c} \right) \pi a^2 Q_\text{pr}.
\label{EQN:f_r}
\end{equation}
If the dust grain has internal density $\rho_\text{gr}$, we can also express
the radial force in terms of the grain mass $m$ and radial velocity $v_r$ as
\begin{equation}
f_\text{pr} = m \frac{\diff v_r}{\diff t} = \frac{4 \pi}{3} \rho_\text{gr} a^3 v_r \frac{\diff v_r}{\diff r},
\end{equation}
where in this problem we neglect the drag force and other external forces.
Equating these two expressions,
\begin{equation}
2 v_r \diff v_r = \left( \frac{3 L Q_\text{pr}}{8 \pi c \rho_\text{gr} a} \right) \left( \frac{\diff r}{r^2} \right).
\end{equation}
Integrating for a dust grain starting at rest at $r = r_0$, one obtains a
relation between dust grain radial velocity and radial distance,
\begin{equation}
v_r = \frac{\diff r}{\diff t} = \sqrt{\frac{3 L Q_\text{pr}}{8 \pi c \rho_\text{gr} a} \left( \frac{1}{r_0} - \frac{1}{r} \right)}.
\label{EQN:drdt_analytic}
\end{equation}
As $r \to \infty$, the dust grain approaches a constant terminal velocity.
Grains starting at smaller initial radii, closer to the radiation source, have
larger terminal velocities.

We place a monochromatic source of luminosity $L = 10^6 \, \text{L}_\odot$ at
the center of a three-dimensional box of side length $160 \, \text{pc}$.  The
volume is tessellated by $256^3$ gas cells with uniform density $n = 1 \,
\text{cm}^{-3}$.  The positions of mesh-generating points are initially
arranged on an equispaced Cartesian lattice.  Then, points are uniformly
randomly displaced in each dimension by up to 20 per cent of the initial cell
size in order to produce an irregular mesh.  This approximates the cell
geometries seen in typical simulations \citep{Vogelsberger2012, Kannan2019a}.
To model the luminosity source, during every time-step photons are injected
in equal amounts into gas cells within $1 \, \text{pc}$ of the box center.  For
completeness, we assume photons have energy $13.6 \, \text{keV}$, though this
choice does not affect our results.

\begin{figure}
\centering
\includegraphics{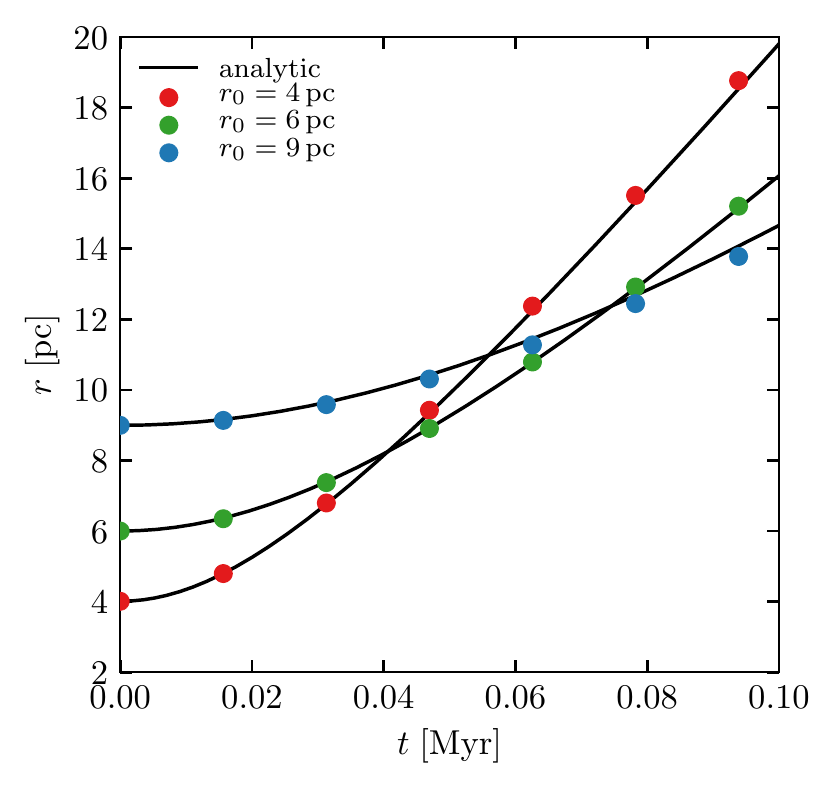}
\caption{Radial distance versus time for a sample of dust grains moving away
from a source of constant luminosity.  Simulation results are shown in colour,
while analytic predictions are in black.  Grains have fixed size but different
values of $r_0$, the initial radial distance from the radiation source.  In
this test, we neglect drag between dust and gas and include only radiation
pressure on dust.}
\label{FIG:radiation_sphere_evolution}
\end{figure}

We place $256^3$ dust particles on an equispaced Cartesian lattice.  Dust
particles are given equal mass to model an initially uniform dust distribution.
The dust particle mass is chosen so that the total dust-to-gas ratio is $0.5$.
However, to focus solely on the effect of radiation pressure, in this test we
neglect any drag coupling between dust and gas.  As a result, the value of the
dust-to-gas ratio does not affect results.  We also assume that only dust
particles provide opacity to radiation: the opacity of gas cells is set to
zero, and we turn off gas thermochemistry so that the gas begins and remains
purely neutral hydrogen.  Thus, the gas is analytically expected to remain at
rest.

Dust particles are assumed to have grains of size $a = 0.01 \, \mu\text{m}$ and
internal density $\rho_\text{gr} = 2.4 \, \text{g} \, \text{cm}^{-3}$.  For
simplicity, we assume that a grain's radiation pressure cross section equals
its geometric cross section and set $Q_\text{pr} = 1$.  When calculating
radiation transport, we adopt the reduced speed of light approximation
\citep[in the notation of][we set $\tilde{c} / c = 1/25$]{Kannan2019a}.  This
reduces the rate at which photons cross the simulation domain and in turn
lessens noise in photon fluxes near the box edges where photons meet.

\begin{figure}
\centering
\includegraphics{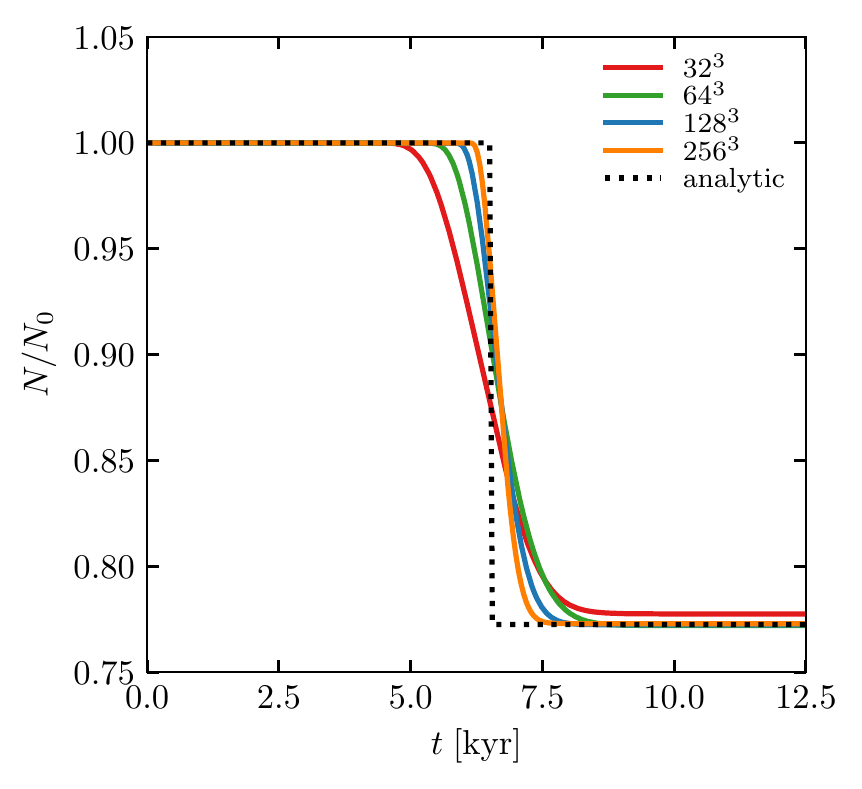}
\caption{Number of photons $N$ remaining as a function of time for a test in
which a burst of radiation is incident on a thin layer of dust.  We normalize
by the initial number of photons $N_0$.  Coloured lines show simulation results
at different resolutions, with the labels indicating number of gas cells.  The
dotted line shows the expected analytic behaviour, with the photon fraction
dropping as it passes through the dust layer.}
\label{FIG:radiation_layer_evolution}
\end{figure}

Figure~\ref{FIG:radiation_sphere_evolution} shows the time evolution of a
sample of dust particles as they move away from the radiation source.  We
select dust particles that begin at initial radial distances of $r_0 = 4, 6, \,
\text{and} \, 9 \, \text{pc}$.  We compare the radial motion of these
simulation particles over a time period of $0.1 \, \text{Myr}$ with the
expected behaviour obtained by numerically integrating
equation~(\ref{EQN:drdt_analytic}).  The initial radiation force is greater for
dust grains closer to the luminosity source.  As a result, by $t = 0.1 \,
\text{Myr}$ the grain initially closest to the luminosity source ($r_0 = 4 \,
\text{pc}$) develops the largest velocity and is at the furthest radial
distance.  In contrast, the grain starting at $r_0 = 9 \, \text{pc}$ lags
behind those two starting closer to the source.  While radiation pressure does
drive the dust outwards, the resulting distribution does not preserve the
initial ordering of grains in terms of radial distance from the luminosity
source.

The simulation results largely agree with the expected behaviour.  There are
some minor deviations above (e.g.~$r_0 = 4 \, \text{pc}$) and below (e.g.~$r_0
= 9 \, \text{pc}$) the analytic solutions, likely influenced by the fact that
we update dust particle positions using kernel-interpolated estimates of the
gas cell radiation flux.  However, we have verified that these deviations
diminish as the resolution of the test problem increases.
Figure~\ref{FIG:radiation_sphere_evolution} demonstrates the suitability of our
hybrid gas cell and dust particle approach for modelling radiation pressure on
dust grains.

\subsection{Photon absorption by dust grains}
We demonstrate the ability of dust grains to absorb photons in our hybrid dust
particle and gas cell scheme using a simplified test problem, where radiation
is incident on a thin layer of dust.

We begin with a lattice of $N^3$ equispaced gas cells of uniform density $n = 1
\, \text{cm}^{-3}$ in a box of side length $L = 160 \, \text{pc}$.  Cell
centers are uniformly randomly displaced by up to 20 per cent in each
coordinate, except for the two leftmost and two rightmost layers of cells along
the $x$ axis, where the mesh remains Cartesian.  A layer of $N^2$ equispaced,
equal-mass dust particles is placed halfway through the box at $x = L/2$, with
the dust particle mass $m_\text{d}$ chosen so that the total dust-to-gas ratio
is $10^{-3}$.  Both dust and gas components start at rest, and in this test of
radiation dynamics we switch off self-gravity and the drag force.  Dust
particles are assumed to consist entirely of $a = 0.01 \, \mu\text{m}$ sized
grains of density $\rho_\text{gr} = 2.4 \, \text{g} \, \text{cm}^{-3}$, and we
take $Q_\text{abs} = 1$.  This simplifies the expected analytic behaviour, but
in principle dust particles could be populated with a full distribution of
grain sizes.

For the initial conditions, we place radiation of uniform energy density in the
$N^2$ gas cells bordering the $x = 0$ edge of the box.  The radiation flux
points in the positive $x$ direction and has its maximally allowed magnitude
\citep[following the notation of][$|\vec{F_r}| = \tilde{c} E_r$]{Kannan2019a}.
We use reduced speed of light $\tilde{c} = c/25$ for photon propagation.  In
this test, we use one UV bin and switch off hydrogen and helium thermochemistry
so that dust grains provide the only source of opacity.  Dust particles spread
their opacity over the $N_\text{ngb} = 64$ neighboring gas cells, though our
results are largely insensitive to this choice.

\begin{figure}
\centering
\includegraphics{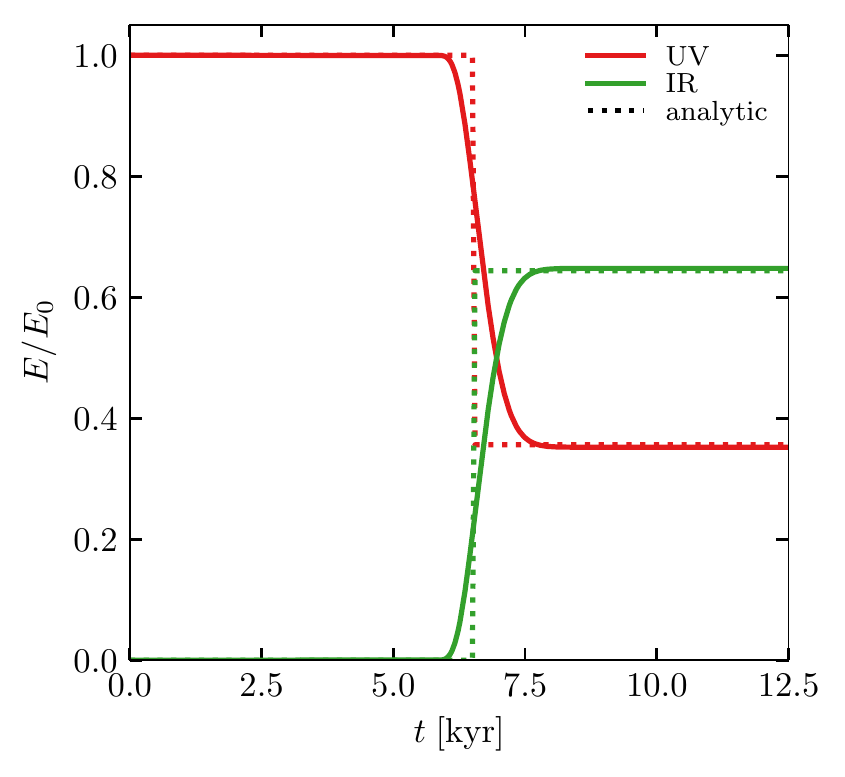}
\caption{Fraction of energy in UV (red) and IR (green) photons in a test where
UV radiation is incident on a thin layer of dust.  Dust grains absorb UV
photons and reemit in the IR, and we turn off absorption of IR photons so that
the total energy in radiation is constant.  Dotted lines show the expected
analytic result, given the optical depth $\tau \approx 1$.}
\label{FIG:radiation_multifrequency_evolution}
\end{figure}

Figure~\ref{FIG:radiation_layer_evolution} shows the number fraction of photons
remaining in the box as a function of time, for four resolution tests ($N =
32, 64, 128, \, \text{and} \, 256$).  Analytically, the photons reach the layer of
dust at $x = L/2$ in time $t = L / (2 \tilde{c}) \approx 6.5 \, \text{kyr}$,
after which the number fraction should drop from $1$ to $\exp(-\tau)$, in terms
of the optical depth
\begin{equation}
\tau = \int_{0}^{L} n(x) \, \sigma \, \diff x
\end{equation}
where $n(x) = \delta(x - L/2) \times N^2 / L^2$ is the number density of dust
particles and $\sigma = (3 m_\text{d}) / (4 \rho_\text{gr} a)$ is the total
cross section of each dust particle.  For our choice of parameters above, $\tau
\approx 0.26$.  There is strong agreement between the analytic prediction and
simulation results, with the late-time photon number fraction dropping to
$\exp(-\tau)$ as expected.  Increasing the resolution of the test produces a
sharper drop in photon number fraction, since the gas cells surrounding the
dust layer have smaller extent.

\subsection{Dust reprocessing}

To demonstrate the ability for dust grains to absorb UV photons and reemit in
the IR, we extend the test presented in
Figure~\ref{FIG:radiation_layer_evolution}, where UV photons incident upon a
thin dust layer were strictly absorbed and not reemitted.  In this new test, we
adopt one UV and one IR radiation bin and convert absorbed UV energy to IR
energy.  We neglect dust grain emission and absorption in the IR, so that the
total energy in radiation is constant and simply shifts between frequency bins.

The initial conditions are the same as in
Figure~\ref{FIG:radiation_layer_evolution}, except that we decrease the fixed
grain size to $a = 2.5 \, \times \, 10^{-3} \, \mu\text{m}$.  Since the dust
particle mass is unchanged, the cross section per dust particle increases by a
factor of four, and the optical depth through the dust layer is $\tau \approx
1$.  We pick a resolution of $N = 128$ for gas cells and dust particles.

\begin{figure}
\centering
\includegraphics{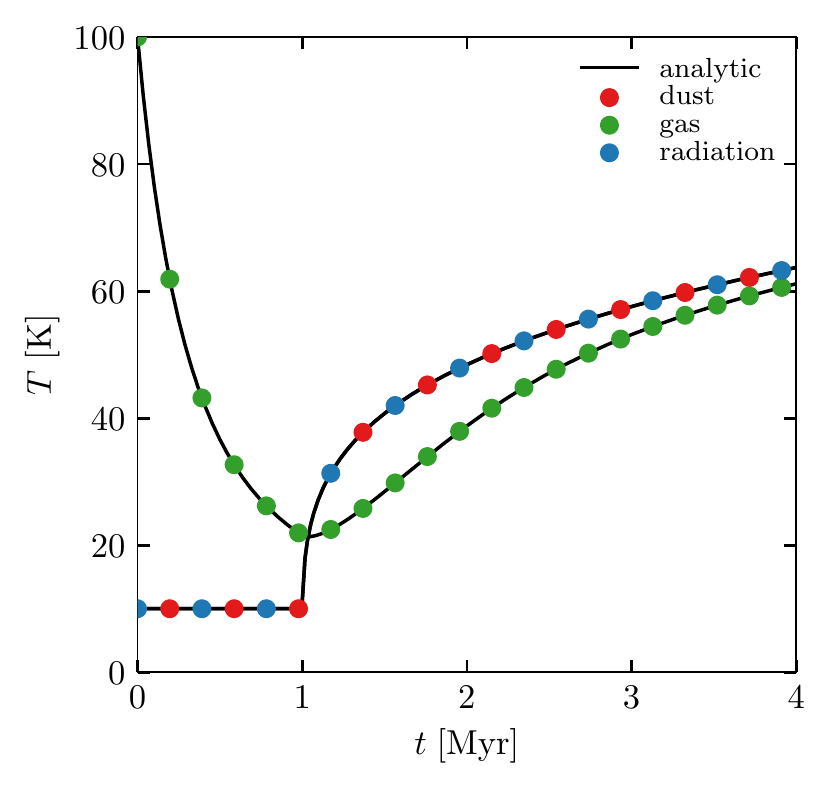}
\caption{Temperature evolution for a test in which dust (red) is coupled to gas
(green) through collisional energy exchange and to IR radiation (blue) through
thermal IR emission and absorption.  The test includes injecting IR photons
into every gas cell at a constant rate starting at $t = 1 \, \text{Myr}$ to
mimic a radiation source.  Coloured points show simulation results, while black
lines indicate the expected solution by numerically integrating the analytic
energy exchange equations.  Dust and radiation temperatures are highly coupled,
while the gas temperature is slower to change.}
\label{FIG:thermal_coupling_evolution}
\end{figure}

\begin{figure*}
\centering
\includegraphics{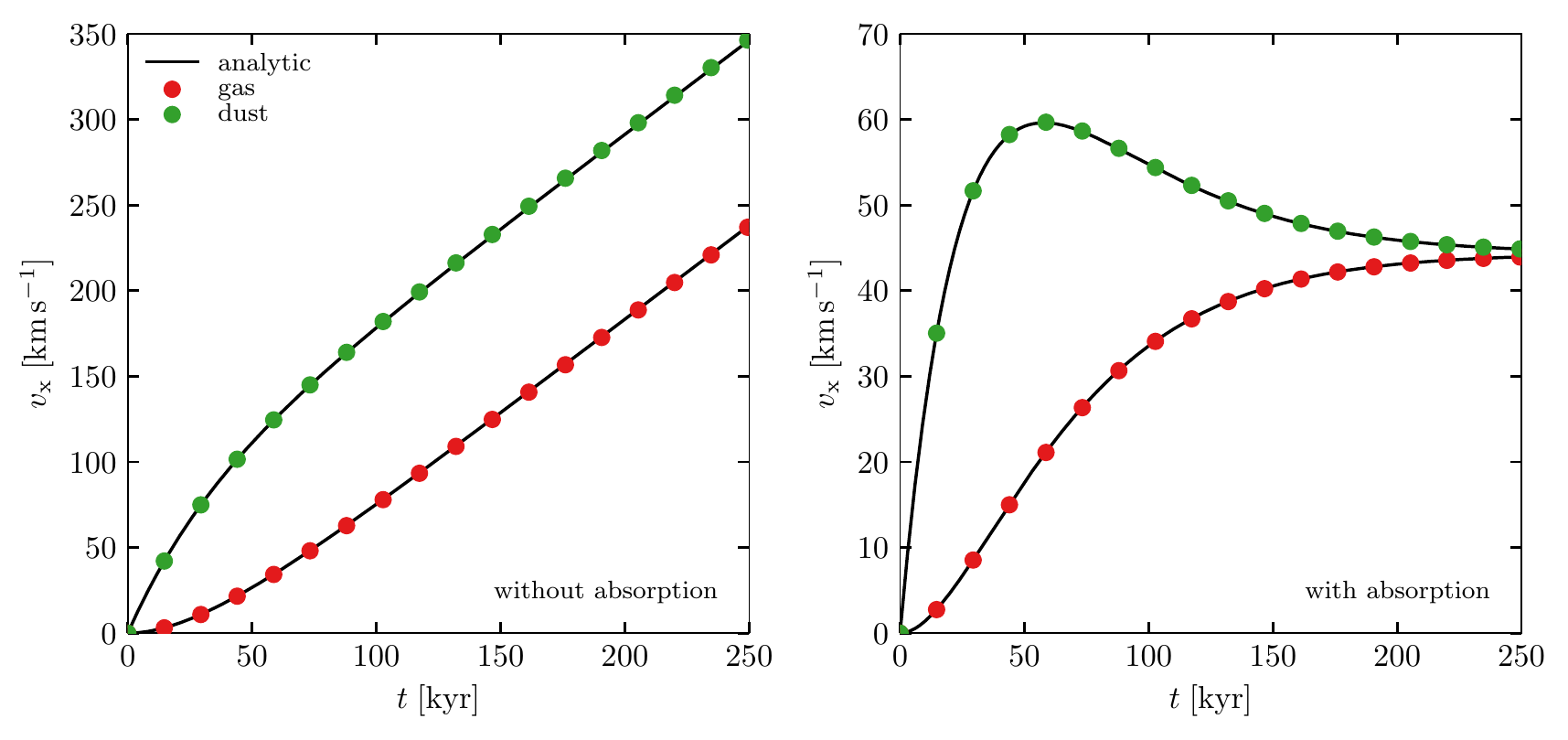}
\caption{Left panel: mean velocity of gas (red) and dust (green) versus time
for a test in which dust grains feel radiation pressure but do not absorb
photons so that the radiation flux is constant.  Analytic profiles are shown in
black.  Radiation accelerates dust, which in turn drags gas.  The rate of
momentum injection from radiation remains constant.  Right panel: similar
results for a test with photon absorption included.  As the photon flux drops,
dust and gas velocities equilibrate through drag.  In both panels, simulation
output is subsampled to improve readability.}
\label{FIG:radiation_momentum_evolution}
\end{figure*}

Figure~\ref{FIG:radiation_multifrequency_evolution} shows the fraction of
radiation energy in UV and IR bins as a function of time, along with the
expected analytic behaviour.  The optical depth $\tau \approx 1$ is chosen in
this test so that a majority of UV energy converts to IR as photons pass
through the dust layer.  We note that, since IR photons are lower energy than
UV photons, this test implies an increase in the total number of photons.  Our
results agree with the analytic prediction that the fraction of energy in UV
and IR bins reaches $e^{-\tau}$ and $1-e^{-\tau}$, respectively.

\subsection{Thermal coupling}

We next present a test in which dust, gas, and radiation are thermally coupled
through the system in equation~(\ref{EQN:thermal_coupling_system}).  In a box
of side length $1 \, \text{kpc}$, we place an irregular mesh of $32^3$ gas
cells with uniform density $n = 1 \, \text{cm}^{-3}$ and an equispaced lattice
of $32^3$ dust particles.  We set the dust-to-gas ratio $D = 1$, fixed grain
size $a = 0.005 \, \mu\text{m}$ (i.e.~here, we do not place grains in multiple
size bins), efficiencies $Q_\text{abs} = 1$, and internal grain density
$\rho_\text{gr} = 2.4 \, \text{g} \, \text{cm}^{-3}$.  Dust and gas components
initially start at rest, although we do not focus on dynamics in this test.

The initial dust and gas temperatures are $T_\text{d} = 10 \, \text{K}$ and
$T_\text{g} = 100 \, \text{K}$, respectively, and we choose the initial IR
radiation density so that the radiation temperature $T_\text{r} \equiv
(E_\text{IR} / a_\text{B})^{1/4} = 10 \, \text{K}$ at the start.  We do not use
a reduced speed of light, so that $\tilde{c} = c$.  To mimic a radiation source
that emits in the IR, starting at $t = 1 \, \text{Myr}$ we inject IR photons
into gas cells at the constant rate $\dot{E_\text{IR}} = 1.2 \times 10^{57} \,
\text{erg} \, \text{Myr}^{-1} \, \text{kpc}^{-3}$.  In this setup, we set
radiation in non-IR bins to zero and do not include other sources of gas
thermal change (e.g.~from photoheating) beyond dust-gas collisional energy
exchange.

Figure~\ref{FIG:thermal_coupling_evolution} shows the evolution of dust, gas,
and radiation temperatures, computed by averaging over simulation gas cells and
dust particles.  Since the gas temperature initially is greater than the dust
temperature, the gas begins to collisionally cool.  Because of the strong
dependence of dust-radiation energy exchange on dust temperature, the dust
temperature does not significantly change as the gas temperature drops.  When
the IR radiation injection begins at $t = 1 \, \text{Myr}$, dust remains highly
coupled with radiation.  In contrast, the gas temperature lags behind the
increase in dust and radiation temperatures.  Our simulation results closely
follow the predictions obtained by integrating the analytic system in
equation~(\ref{EQN:thermal_coupling_system}), where we adopt the silicate grain
heat capacity from \citet{Dwek1986} to relate dust internal energy change to
dust temperature change.  This test shows how dust, gas, and radiation
temperatures can evolve according to energy exchange processes and external
sources.

\subsection{Coevolution of radiation, dust, and gas through radiation pressure and drag}\label{SEC:drag_radiation}

Previous tests illustrated the effect of drag without radiation
(e.g.~Figure~\ref{FIG:dustybox_backreaction_evolution}) or radiation without
drag (e.g.~Figures~\ref{FIG:radiation_sphere_evolution},
\ref{FIG:radiation_layer_evolution},
\ref{FIG:radiation_multifrequency_evolution},
and~\ref{FIG:thermal_coupling_evolution}).  We now demonstrate the ability of
our methods to simultaneously handle drag and radiation.

The initial conditions consist of uniform dust and gas distributions at rest
and are identical to those used in Section~\ref{SEC:thermal_coupling}, except
that we set the dust-to-gas ratio $D = 0.5$ and fixed grain size $a = 0.1 \,
\mu\text{m}$.  We initialise a uniform monochromatic radiation field whose flux
has magnitude $F_0 = 10 \, \text{erg} \, \text{s}^{-1} \, \text{cm}^{-2}$
pointing in the $\vechat{x}$ direction.  Dust particles contribute the only
opacity to radiation (i.e.~we neglect gas thermochemistry and opacity
contributed by gas cells), and photons propagate at the reduced speed of light
$\tilde{c} = c / 1000$.  As a result, in this test dust grains accelerate from
photon radiation pressure and in turn accelerate the gas through drag.  In
addition to our fiducial setup where dust grains both feel radiation pressure
and absorb photons, we also perform a test where photon absorption is turned
off so that analytically the radiation flux $\vec{F}(t) = F_0 \vechat{x}$ at
all times.

To ensure our simulation results can be compared with analytic predictions, we
require that the stopping time-scale $t_\text{s}$ is constant.  For this test
only, we disable the injection of thermal energy into the gas from drag
dissipation of kinetic energy.  Thus, the sound speed is expected to remain
constant.  Additionally, in equation~(\ref{EQN:t_s}) we neglect the correction
term in parentheses.  In Appendix~\ref{SEC:appendix_analytic}, we show how to
analytically integrate the dust and gas equations of motion.

The choice of parameters above yield a stopping time-scale $t_\text{s}$ and
photon decay time-scale $t_\text{d}$ both on the order of $40 \, \text{kyr}$.
If we had $t_\text{s} \ll t_\text{d}$, radiation pressure would essentially
move dust and gas together as one, while $t_\text{s} \gg t_\text{d}$ would see
photons rapidly deposit momentum into dust and decay away before drag
accelerated gas.  In our setup, dust, gas, and radiation coevolve over similar
time-scales.

Figure~\ref{FIG:radiation_momentum_evolution} shows the mean velocity of gas
cells and dust particles as a function of time.  In addition to the fiducial
setup with $t_\text{s} \approx t_\text{d}$, we also perform a test neglecting
photon absorption by dust grains.  In this test, which is equivalent to taking
$t_\text{d} \to \infty$, radiation flux remains constant and continually
imparts momentum to the dust at a fixed rate.  Runs with and without photon
absorption show different qualitative evolution, though in both cases
simulation results closely follow analytic results.

Without absorption, dust and gas velocities increase in time without bound.
Initially, dust accelerates more quickly than gas, since only dust feels
radiation pressure.  At later times, dust and gas feel roughly the same
acceleration, with the velocity offset $v_\text{d} - v_\text{g} \approx 108 \,
\text{km} \, \text{s}^{-1}$ matching the analytic prediction of $F_0 \kappa
t_\text{s} / c$.

With absorption, dust still feels a larger initial acceleration.  However, as
photons get absorbed, the rate of momentum injection decays to zero and dust
and gas velocities equilibrate via the drag force.  Thus, there is no long-term
velocity offset between dust and gas components.  The steady-state velocities
$v_\text{d} \approx v_\text{g} \approx 44 \, \text{km} \, \text{s}^{-1}$ agree
with the analytic expectation of $F_0 \kappa D t_\text{d} / [(1 + D) c]$.

\section{Conclusions}\label{SEC:conclusion}


We have developed a framework to couple dust physics and radiation
hydrodynamics in the moving-mesh code \textsc{arepo-rt}, where dust is modelled
using live simulation particles and radiation is handled on an unstructured
mesh.  We first extend our implementation of the aerodynamic drag force that
couples dust and gas motion to properly capture dynamics at arbitrary
dust-to-gas ratios, even those well above Galactic values.  We then describe
kernel-smoothed methods to model the interaction between dust and radiation,
where grain cross sections from simulation dust particles are interpolated onto
neighboring gas cells to affect radiation fields in a self-consistent manner.

We detail our approach for calculating radiation pressure forces on populations
of dust grains, including those covering a range of different grain sizes.
This method is tested using a constant luminosity source in a medium with an
initially uniform dust density, for which analytic behaviour is known.  We
model absorption of photons by local dust grains and maintain the ability of
\textsc{arepo-rt} to treat multifrequency radiation, where UV and optical
photons are single-scattered and IR photons are multi-scattered.  We illustrate
the ability for dust to absorb photons passing through a thin layer of dust of
known optical depth.  Instead of assuming dust and gas are in local
thermodynamic equilibrium, we allow dust and gas to have separate temperatures
and model their energy exchange through collisional processes.  Dust and
radiation also exchange energy through absorption and thermal dust emission at
IR wavelengths.

Finally, we demonstrate the evolution of a mixture of dust, gas, and radiation
using our hybrid dust particle and gas cell scheme.  Dust and gas are coupled
through aerodynamic drag, and dust and radiation are coupled through radiation
pressure and photon absorption.  In agreement with analytic predictions, we
verify that dust and gas velocities initially deviate as photons accelerate
dust but later equilibrate through drag as all photons get absorbed.  This
behaviour more closely mirrors the expected physics compared to models that
assume a perfect coupling between dust and gas.

We emphasise the original nature of our methods in comparison to typical
radiation hydrodynamics schemes that model dust opacity using fixed values or
simplified functions of hydrodynamic quantities like temperature.  By treating
dust using live simulation particles subject to dynamical forces, our approach
avoids the need to make assumptions about the local dust-to-gas ratio and dust
distribution and instead calculates dust opacity self-consistently based on the
local grain abundance.  While there are codes to perform radiative transfer in
arbitrary dusty media, these usually run in post-processing and not
concurrently with hydrodynamic evolution.  As radiation hydrodynamic
simulations of galaxy formation become more computationally tractable, our
model for the interaction of dust and radiation is an alternative to
simulations that neglect a direct treatment of dust.

\section*{Acknowledgements}

We thank Volker Springel for sharing access to \textsc{arepo}.  MV acknowledges support through an MIT RSC award, a Kavli Research Investment Fund, NASA ATP grant NNX17AG29G, and NSF grants AST-1814053, AST-1814259 and AST-1909831. PT acknowledges support from NSF grant AST-1909933. The simulations
were performed on the joint MIT-Harvard computing cluster supported by MKI and
FAS.  RM acknowledges support from the DOE CSGF under grant number
DE-FG02-97ER25308. 

\bibliographystyle{mn2e}
\bibliography{bibliography}

\appendix

\section{Analytic solution for radiation, dust, and gas coevolution test problem}\label{SEC:appendix_analytic}

We consider the analytic evolution of the dust and gas mixture described in
Section~\ref{SEC:drag_radiation} according to drag and radiation pressure.

By symmetry, the $x$ components of the dust and gas velocities, $v_\text{d}$
and $v_\text{g}$ respectively, are independent of position and dependent only
on time.  They evolve according to
\begin{equation}
\frac{\diff v_\text{d}}{\diff t} = -\frac{v_\text{d}(t) - v_\text{g}(t)}{t_\text{s} (1 + D)} + \frac{F(t) \kappa}{c}
\label{EQN:dvd_dt_app}
\end{equation}
and
\begin{equation}
\frac{\diff v_\text{g}}{\diff t} = D \frac{v_\text{d}(t) - v_\text{g}(t)}{t_\text{s} (1 + D)},
\label{EQN:dvg_dt_app}
\end{equation}
where $F(t)$ is the radiation flux in the $\vechat{x}$ direction, $D =
\rho_\text{d} / \rho_\text{g}$ is the dust-to-gas ratio, $t_\text{s}$ is the
drag stopping time-scale  and $c$ is the speed of light.  Here, the grain
opacity $\kappa = (3 Q_\text{abs}) / (4 \rho_\text{gr} a)$ for grain size $a$
and grain density $\rho_\text{gr}$ is the absorption cross section \textit{per
unit dust mass}.  Additionally, the gas remains at uniform density and internal
energy and feels no pressure gradient.

To solve these equations analytically, we first define the state vector
\begin{equation}
\vec{w}(t) =
\begin{pmatrix}
v_\text{d}(t) \\
v_\text{g}(t) \\
\end{pmatrix},
\end{equation}
and the matrix
\begin{equation}
\vec{A} =
\frac{1}{t_\text{s} (1 + D)}
\begin{pmatrix}
-1 & 1 \\
D & -D \\
\end{pmatrix}.
\end{equation}
Matrix $\vec{A}$ has eigenvalues $\lambda_1 = -1 / t_\text{s}$ and
$\lambda_2 = 0$, with corresponding eigenvectors $\vec{x_1} = (-1/D,
1)^\text{T}$ and $\vec{x_2} = (1, 1)^\text{T}$, respectively.  We introduce the
inhomogeneous source vector
\begin{equation}
\vec{f}(t) =
\begin{pmatrix}
F(t) \kappa / c \\
0 \\
\end{pmatrix}.
\label{EQN:source_flux}
\end{equation}
Over elapsed time $t$, photons experience an optical depth of $\tau(t) =
\rho_\text{d} \kappa \tilde{c} t$, where $\tilde{c}$ is the (possibly reduced)
speed of photon propagation.  This gives a uniform radiation flux at time $t$
of $F(t) = F_0 \exp(-\tau(t)) = F_0 \exp(-t / t_\text{d})$, written in terms of
decay time-scale $t_\text{d} \equiv 1 / (\rho_\text{d} \kappa \tilde{c})$.  The
coupled equations~(\ref{EQN:dvd_dt_app}) and~(\ref{EQN:dvg_dt_app}) can be
written in vector form as
\begin{equation}
\frac{\diff \vec{w}}{\diff t} = \vec{A} \vec{w}(t) + \vec{f}(t).
\label{EQN:dw_dt}
\end{equation}
Since gas and dust start at rest, the initial condition is $\vec{w}(0) = (0,
0)^\text{T}$.

We define the helper matrix
\begin{equation}
\vec{X}(t) =
\begin{pmatrix}
\vec{x_1} e^{\lambda_1 t} & \vec{x_2} e^{\lambda_2 t} \\
\end{pmatrix}
=
\begin{pmatrix}
-\frac{1}{D} e^{-t/t_\text{s}} & 1 \\
e^{-t/t_\text{s}} & 1 \\
\end{pmatrix},
\end{equation}
whose inverse is given by
\begin{equation}
\vec{X}^{-1}(t) =
\frac{1}{1+D}
\begin{pmatrix}
-D e^{t/t_\text{s}} & D e^{t/t_\text{s}} \\
D & 1 \\
\end{pmatrix}.
\end{equation}
Then, the general solution to equation~(\ref{EQN:dw_dt}) is
\begin{equation}
\vec{w}(t) = C_1
\begin{pmatrix}
-1/D \\
1 \\
\end{pmatrix}
e^{-t/t_\text{s}}
+ C_2
\begin{pmatrix}
1 \\
1 \\
\end{pmatrix}
+ \vec{w_p}(t)
\end{equation}
where $C_1$ and $C_2$ are constants necessary to match the initial condition
and
\begin{equation}
\vec{w_p}(t) = \vec{X}(t) \int \vec{X}^{-1}(t) \vec{f}(t) \, \diff t
\label{EQN:w_p}
\end{equation}
is the particular solution.  Since the only time dependence in
equation~(\ref{EQN:w_p}) is exponential, the particular solution can be
directly computed.  For brevity, we omit the form of $\vec{w_p}(t)$ and
constants $C_1$ and $C_2$.

By defining the time-scale $t_*$ via
\begin{equation}
\frac{1}{t_*} \equiv \frac{1}{t_\text{s}} - \frac{1}{t_\text{d}},
\end{equation}
we can write the analytic velocity evolution for dust as
\begin{equation}
\begin{split}
v_\text{d}(t) = \frac{F_0 \kappa}{(1+D) c}  \bigg[ &- t_* e^{-t/t_\text{s}} + D t_\text{d} \\
&+ t_* e^{-t/t_\text{d}} - D t_\text{d} e^{-t/t_\text{d}} \bigg]
\end{split}
\end{equation}
and for gas as
\begin{equation}
\begin{split}
v_\text{g}(t) = \frac{D F_0 \kappa}{(1+D) c}  \bigg[ & t_* e^{-t/t_\text{s}} + t_\text{d} \\
&- t_* e^{-t/t_\text{d}} - t_\text{d} e^{-t/t_\text{d}} \bigg].
\end{split}
\end{equation}
Analytic profiles for the case where dust grains feel radiation pressure but do
not absorb photons (i.e.~substituting $F(t) = F_0$ in
equation~\ref{EQN:source_flux}) can be obtained by taking the limit as the
photon decay time-scale $t_\text{d} \to \infty$.

\label{lastpage}

\end{document}